\documentclass[prl,showpacs,amsmath,amssymb,twocolumn,aps]{revtex4}

\usepackage{graphicx,graphics}
\usepackage{epsfig}
\usepackage{dcolumn}
\usepackage{bm}
\usepackage{natbib}
\usepackage{times}

\newcommand{\dal}{\ensuremath{\Delta \alpha/ \alpha}}

\newcommand{\beq}{\begin{equation}}
\newcommand{\eeq}{\end{equation}}

\begin{document} 
\title{Conjugate 18cm OH Satellite Lines at a Cosmological Distance}

\author{Nissim Kanekar} 
\email{nissim@astro.rug.nl}
\affiliation{Kapteyn Astronomical Institute, University of Groningen, The Netherlands,}
\author{Jayaram N. Chengalur}
\email{chengalur@ncra.tifr.res.in}
\affiliation{National Centre for Radio Astrophysics, Pune 411 007, India }
\author{Tapasi Ghosh}
\email{tghosh@naic.edu}
\affiliation{Arecibo Observatory, Arecibo, PR 00612, USA}

\date{\today}

\begin{abstract}
We have detected the two 18cm OH satellite lines from the $z \sim 0.247$ 
source PKS1413+135, the 1720~MHz line in emission and the 1612~MHz line in 
absorption. The 1720~MHz luminosity is $L_{\rm OH} \sim 354 
L_\odot$, more than an order of magnitude larger than that of any other 
known 1720~MHz maser. The profiles of the two satellite lines are conjugate, 
implying that they arise in the same gas. This allows us to test for any 
changes in the values of fundamental constants, without being affected by 
systematic uncertainties arising from relative motions between the gas clouds 
in which the different lines arise. Our data constrain changes in $G \equiv g_p 
[ \alpha^2/y]^{1.849}$, where $ y \equiv m_e/m_p$; we find $\Delta G/G = 2.2 \pm 3.8 
\times 10^{-5}$, consistent with no changes in $\alpha$, $g_p$ and $y$.
\end{abstract}

\maketitle 
\section{Introduction}

In recent times, much interest has centred on the possibility that 
fundamental ``constants'' such as the fine structure constant $\alpha$ 
might vary with cosmic time (e.g.~\cite{webb99,webb01,srianand04}). 
Many theoretical models, including Kaluza-Klein theories and 
superstring theories, predict spatio-temporal variation of these 
constants. However, terrestrial experiments have so far shown no evidence 
for such changes. For example, the strongest presently available 
constraints on changes in $\alpha$ arise from isotopic abundances measured
in the Oklo natural fission reactor, which give the constraint 
$\dal < 1.2 \times 10^{-7}$ on fractional changes in the fine 
structure constant \cite{damour96}. Uzan \cite{uzan03} provides an 
excellent review of current experimental and observational constraints 
on changes in $\alpha$ and other fundamental constants.

While terrestrial studies such as the Oklo experiment have a high 
sensitivity, these measurements only probe a small fraction of the 
age of the Universe. For example, the Oklo reactor operated about 
$1.8$~Gyr ago, less than a sixth of the age of the Universe 
\footnote{Throughout this paper, we use an LCDM cosmology,
with $\Omega_m = 0.3$, $\Omega_\Lambda = 0.7$ and $H_0 = 70$~km/s~Mpc$^{-1}$.}. 
Thus, current 
terrestrial experiments cannot rule out earlier changes in the fundamental 
constants, quite possible if these changes are non-monotonic in nature. 

Astrophysical studies of redshifted spectral lines provide a powerful 
probe of putative changes in fundamental constants over large fractions
of the age of the Universe (e.g.~\cite{webb99,carilli00,chengalur03,srianand04}).
Perhaps the most interesting of these estimates arise 
from the new ``many-multiplet'' method \cite{dzuba99,webb99}, which has 
been used in conjunction with Keck telescope spectra to obtain $\dal = 
(-0.54 \pm 0.12) \times 10^{-5}$ over the redshift range $0.2 < z < 3.7$ 
\citep{murphy03} (see, however, \cite{bekenstein03}). On the other hand,
the many-multiplet method was recently applied to Very Large Telescope (VLT) 
data to obtain $\dal = (-0.6 \pm 0.6) \times 10^{-6}$ over the redshift range 
$0.4 < z < 2.3$ \cite{srianand04}, in direct conflict with 
the Keck results. Further, many theoretical analyses predict that changes in 
$\alpha$ should be accompanied by much larger changes in other constants, 
such as the ratio of electron mass to proton mass $m_e/m_p$ (e.g. 
\cite{calmet02,langacker02}).  However, fractional changes in $m_e/m_p$ 
have been found to be less than $(-0.5 \pm 3.6) \times 10^{-5}$ over a similar 
redshift range  ($ 0< z <2.75$) \cite{ubachs04}. 

Given the importance of changes in fundamental constants for theoretical 
physics, it is important that this be tested by independent {\it techniques}.
This is especially true since the same technique (i.e. the many-multiplet method) 
has yielded strongly divergent results when applied to data from different 
telescopes, suggesting that systematic errors play a significant role (e.g. 
\cite{ashenfelter04}). We have 
recently (\cite{chengalur03}, hereafter Paper I; see also \cite{darling03})  
developed a new technique to simultaneously constrain changes 
in $\alpha$, the ratio of electron mass to proton mass $y \equiv m_e/m_p$ 
and the proton gyromagnetic ratio $g_p$, using the multiple 18cm OH lines. 
This makes use of the fact that these lines arise from two very different 
physical mechanisms, Lambda-doubling and hyperfine splitting \cite{townes55}, 
and thus have very different dependences on $\alpha$, $y$ and $g_p$. If $g_p$ 
is assumed to remain unchanged (e.g.~\cite{webb01,carilli00}),
observations of all four OH 18cm transitions in a single cosmologically 
distant galaxy can be used to simultaneously estimate any changes in 
$y$ and $\alpha$. Further, since these lines arise from the same species, 
a comparison of the OH column density obtained from each line can 
be used to test whether the lines arise from the same gas. This is 
often an issue in other techniques, since lines from different species are
used in the comparison, giving rise to systematic uncertainties arising 
from the possibility of relative motions between the gas clouds in which 
the different lines arise.

Only the ``main'' 18cm OH lines (at frequencies of $\sim$ 1667 and 1665~MHz) 
have hitherto been detected at cosmological distances 
\cite{chengalur99,kanekar02,kanekar03}; the satellite lines (at $\sim$ 1612 
and 1720~MHz), which are critical to the above test, have not been found until 
now. We report here the first detection of the satellite OH lines at 
cosmologically significant distances, from the $z \sim 0.247$ source PKS1413+135.

\section{Spectra and results}

\begin{figure*}
\centering
\epsfig{file=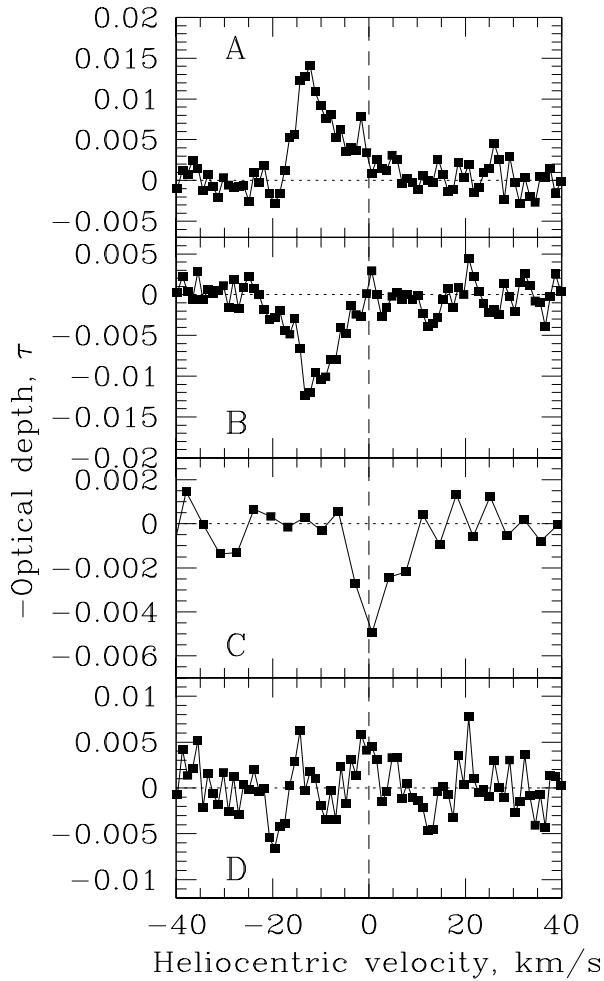,height=5.5in,width=3.3in}
\caption{ OH optical depth $\tau$ v/s heliocentric velocity (relative 
	to $z = 0.24671$). [A] WSRT 1.1~km/s~resolution 1720~MHz spectrum.
	  [B] WSRT 1.1 km/s~resolution 1612~MHz spectrum.
	  [C] GMRT 3.5~km/s~resolution 1667~MHz spectrum.
	  [D] Sum of WSRT 1612 and 1720~MHz spectra. The spectrum is 
	  	consistent with noise. 
	}
\label{fig:spectra}
\end{figure*}

     Spectra of the redshifted OH 1612 and 1720~MHz lines from PKS1413+135 were 
obtained with the Westerbork Synthesis Radio Telescope (WSRT) in June 2003, 
while the 1665 and 1667~MHz transitions were observed with the Giant Metrewave 
Radio Telescope (GMRT) in October 2001. The top three panels of {Fig. 1} show 
the 1720, 1612 and 1667~MHz spectra, respectively, while the bottom panel shows 
the sum of the 1612 and 1720~MHz optical depth profiles; the velocity resolution 
is $\sim 1.1$~km/s for the satellite lines and $\sim 3.5$~km/s for the 1667~MHz 
line. Note that PKS1413+135 is unresolved by the GMRT and WSRT beams; the optical 
depths plotted here are the ratio of line depth to the total flux density for 
the different transitions. The RMS noise in each of the satellite spectra is 
$\sim 0.002$ (per 1.1~km/s channel) while that in the 1667~MHz spectrum is 
$\sim 0.0008$ (per 3.5~km/s channel), in units of optical depth.

The most striking feature of {Fig. 1} is that the 1612 and 1720~MHz lines are 
conjugate to each other; this can be clearly seen in Fig.~\ref{fig:spectra}[D], 
where the sum of the 1612 and 1720~MHz optical depth profiles is consistent with 
noise. We note that the weak feature seen at $\sim -20$~km/s in 
Fig.~\ref{fig:spectra}[D] is not statistically significant ($< 3 \sigma$, 
even after smoothing the profile to increase the signal-to-noise ratio). The 
peaks of the two profiles also arise at the same redshift, within the error bars.

Such conjugate behavior of the satellite lines has been seen earlier in 
extra-galactic sources \cite{langevelde95,vermeulen03} and essentially arises 
due to competition between two decay routes to the {$2\Pi_{3/2}$ (J = 3/2)} ground state,
after the molecules have been pumped from the ground to the higher excited 
rotational states \cite{elitzur92}. The pumping can take place either due to
collisions or far-infrared (FIR) radiation. If the last step of the 
cascade is the ${119\mu}$ transition {$2\Pi_{3/2}$ (J = 5/2)} $\rightarrow$ 
{$2\Pi_{3/2}$ (J = 3/2)}, the 1720~MHz line is inverted and the 1612~MHz line 
anti-inverted; conversely, if the final step is the ${79 \mu}$ decay 
{$2\Pi_{1/2}$ (J = 1/2)} $\rightarrow$ {$2\Pi_{3/2}$ (J = 3/2)}, it results in 
1612~MHz inversion and 1720~MHz anti-inversion \cite{elitzur92}. The fact that 
the 1720~MHz line is seen in emission and the 1612~MHz line in absorption in 
PKS1413+135 implies that the former route dominates 
and gives the constraint $N_{\rm OH}/\Delta V \lesssim 10^{15}$~cm$^{-2}$~(km/s)$^{-1}$
\cite{langevelde95}, where $N_{\rm OH}$ is the OH column density and $\Delta V$,
the velocity width; in our case, $\Delta V \sim 20$~km/s, implying that 
$N_{\rm OH} \lesssim 2 \times 10^{16}$~cm$^{-2}$. Further, all pumping mechanisms
require  that the 119$\mu$ transition be optically thick for the 1720~MHz 
line to be inverted; this corresponds to the lower limit $N_{\rm OH}/\Delta V  \gtrsim  
1 \times 10^{14}$~cm$^{-2}$~(km/s)$^{-1}$ \cite{langevelde95}, i.e. $N_{\rm OH} \gtrsim
2 \times 10^{15}$~cm$^{-2}$ in the present case.  We thus obtain the range $2 \times 
10^{15} \lesssim N_{\rm OH} \lesssim 2 \times 10^{16}$~cm$^{-2}$ for the OH 
column density.  At lower $N_{\rm OH}$, the 119$\mu$ line is optically thin and 
the 1720~MHz line is not inverted; at higher $N_{\rm OH}$, the inversion switches 
to the 1612~MHz line. Implications for other physical conditions in the 
absorbing/emitting cloud will be discussed elsewhere.

The 1720~MHz profile implies an OH luminosity $L_{\rm OH} \sim 354 L_\odot$,
well into the OH megamaser range \cite{baan91}. This is by far the brightest known
1720~MHz maser line, more than an order of magnitude more luminous than Arp~220 
($L_{\rm OH} \sim 12 L_\odot$), the only other known 1720~MHz megamaser \cite{baan87}. 
Interestingly enough, PKS1413+135 is an excellent candidate for {\it main line} OH
megamaser emission, on the basis of its high FIR flux \cite{baan91,darling02}; 
in fact, the empirical relation, ${\rm log} \; L_{\rm OH} = (1.2 \pm 0.1) \: 
{\rm log} \: L_{\rm FIR} - (11.7 \pm 1.2)$ (in solar units) \cite{darling02} 
predicts a main line megamaser luminosity of $L_{\rm OH} \sim  300 L_\odot$, similar
to that obtained in the 1720~MHz line. It is unclear whether 
this is merely a coincidence, especially given the spread in the relation 
between $L_{\rm OH}$ and $L_{\rm FIR}$ \cite{darling02}.

\section{Constraining the variation of fundamental constants}

Crucially, the conjugate behavior of the satellite lines guarantees that {\it they
arise from the same gas}. However, the main line profiles are quite
different, suggesting that this absorption occurs at a different spatial location. The
main lines (the 1665~MHz line is not shown here) peak at $z = 0.24671$,
coincident in velocity with other molecular species (such as CO and HCO$^+$)
that have been detected in PKS1413+135 \cite{wiklind97}. They also show a
``tail'' at positive velocities, with a total velocity spread of $\sim 15$~km/s.
On the other hand, while the two satellite lines have a similar velocity
width ($\sim 20$~km/s), they show blueshifted, edge-brightened profiles, with
the absorption/emission peaking $\sim -15$~km/s~ away from $z = 0.24671$.
Further, while the satellite features are seen to extend to $v = 0$, the main
lines are not detected at the velocity corresponding to the peak of the satellite
lines. 

Most of the astrophysical techniques used to estimate changes in fundamental 
constants (e.g. \cite{carilli00,murphy03,srianand04}) involve a 
comparison between the redshifts of spectral lines of different species, i.e. 
the assumption that these different species have no velocity offsets between them. 
In fact, even a comparison between lines of the same species (e.g. 
\cite{ubachs04,quast04}) does not rule out the above systematic uncertainty, 
as the different lines might well be excited in different regions of the gas 
cloud. In the present case, the very 
different shapes of the profiles of the ``main'' and ``satellite'' line 
profiles implies that one cannot compare the peak redshifts of all four lines 
to simultaneously constrain changes in $y \equiv m_e/m_p$ and $\alpha$. 
However, the conjugate behavior of the OH satellite lines implies that they 
arise from precisely the same gas; we hence have a unique situation where 
{\it systematic velocity offsets between two spectral lines are not an issue}. Again, 
the satellite line frequencies have very different dependences on the 
``constants'' $\alpha$, $y \equiv m_e/m_p$ and the proton g-factor $g_p$; this  
can be clearly seen from equations~(9) and (11) in Paper 1 (note that the 
sum of satellite line frequencies is equal to the sum of main line frequencies 
in the OH radical).  A comparison between the measured 1612 and 1720~MHz redshifts 
thus allows one to measure changes in the quantity $G \equiv g_p 
[ \alpha^2/y]^{1.849}$. In particular, if $z_s$ and $z_d$ are the redshifts 
derived from the sum and difference of the measured 1612 and 1720~MHz line 
frequencies, respectively, it can be shown that ${\Delta{G}/G = [ z_{s}-z_{d}] / 
( 1 + \bar{z})}$, where $\bar{z}$ is the average of $z_{s}$ and $z_{d}$. 

As mentioned earlier, the peaks of the two satellite profiles arise at the 
same redshift, within our observing resolution of $\sim 4.8$~kHz. However, 
since the two profiles are conjugate, we can improve the accuracy of the 
comparison between the peak redshifts by fitting a template profile to each 
spectrum. Note that the same template is fit to both spectra and this procedure 
is carried out solely for the purpose of comparing the peak redshifts.
We use a double-Gaussian template for this purpose, to account for both the 
narrow peak as well as the extended shoulder. Several of these two-component 
fits were found to be statistically indistinguishable; in essence, the profile 
cannot be uniquely decomposed into two Gaussians as the signal-to-noise ratio of 
the shoulder is quite low. Using different templates allowed us to estimate 
the systematic error in the measured redshifts, arising due to the choice of 
template. In all cases, the redshifts of the peak were found to agree to
within $2.8$~kHz, for both the 1720~MHz and 1612~MHz lines. We hence 
use this figure as our estimate of the error in the redshifts (including both 
systematic and random errors). We note that attempts were also made to 
fit a single Gaussian component to the narrow peak in both satellite line 
profiles; again, the peak redshifts were found to be in agreement within 
an error of 2.8~kHz. 

The sum of the frequencies of the peaks of the satellite profiles is then
measured to be $2673.356 \pm 0.004$~MHz, while their difference is 
$86.8733 \pm 0.004$~MHz. In both cases, the errors have been added in 
quadrature -- it should be emphasized that this is a conservative estimate 
in the case of the difference of the frequencies, since the systematic 
error in the choice of the template cancels out here, to first order. The above 
frequencies correspond to redshifts $z_{s} = 0.246658 \pm 0.0000015 $ 
and $z_{d} = 0.24663 \pm 0.000046$, respectively, in agreement within the 
error bars. This gives the limit $\Delta{G}/G = 2.2 \pm 3.8 \times 10^{-5}$ 
on changes in the quantity $G$ from $z \sim 0.247$ to the present epoch.
Again, we emphasize that the errors on $\Delta{G}/G$ are dominated by the 
error on {\it the difference in frequencies} and the latter is certainly 
lower than 2.8~kHz, since the systematic error from the fitting procedure
cancels to first order; we choose, however, to remain conservative in our 
error estimates. 

We note that the strong dependence of the quantity $G$ on the fine 
structure constant $\alpha$ ($ G \propto \alpha^{3.698}$) implies 
that the observations are extremely sensitive to changes in $\alpha$. 
If we assume, as is often done, (e.g. \cite{carilli00}) that neither 
$g_p$ nor $y$ varies with cosmological epoch, the above constraint 
on $\Delta G/G$ translates to $\Delta{\alpha}/\alpha = 0.6 \pm  1.0 \times 
10^{-5}$. Similarly, $G$ also has a strong dependence on the electron-proton 
mass ratio $y \equiv m_e/m_p$ (although weaker than that on $\alpha$), 
with $G \propto y^{-1.849}$; we obtain the constraint $\Delta y/y = (-1.2 \pm 2.0) 
\times 10^{-5}$, assuming that $\alpha$ and $g_p$ remain unchanged. This is 
similar to the best current limits on changes in $y$ from observations of the 
Lyman and Werner band lines of molecular hydrogen \cite{ubachs04}, although the 
latter constraints, of course, probe a larger redshift range. We also 
note that the changes in $\alpha$ and $y$ are of opposite sign and it would 
require something of a conspiracy for both these quantities to vary, while 
leaving $G$ unchanged. The data towards PKS1413+135 are thus consistent with the 
different constants remaining constant from $z \sim 0.247$ to today. 

The relative sensitivity of the present technique adds to its importance. 
While results from the many-multiplet method (e.g. \cite{murphy03,srianand04})
on fractional changes in $\alpha$ have lower errors at the present time, 
this method requires the use of many absorption systems, both to cancel 
systematic effects and to increase sensitivity. Further, as mentioned earlier, 
the conflicting results from analyses based on Keck and VLT spectra suggest 
that unknown systematics dominate the errors here. On the other hand, our 
present limits are derived from a single system and provide an entirely 
independent constraint on changes in $\alpha$, $y$ and $g_p$. It should 
also be noted that the present satellite spectra have a resolution of only 
$\sim 1.1$~km/s; higher resolution spectroscopy in both lines will allow 
far tighter constraints on changes in $G \equiv g_p [ \alpha^2/y ]^{1.849}$. 
For example, observations of PKS1413+135 with the Square Kilometer Array 
\cite{dayton04}, a next generation radio telescope, will be able to 
detect fractional changes $\Delta G / G \sim few \times 10^{-7}$. 
This implies a sensitivity of $\sim 10^{-7}$ to changes in $\dal$, comparable 
to those obtained with the Oklo reactor experiment. 

Finally, the 1720~MHz inversion arises due to an over-population of the 
$F = 2$ level of the ground state relative to the 
$F = 1$ level, as transitions from the $F = 3$ level of the $J = 5/2$ state 
to the latter level are forbidden. This should also give rise to an 
over-population of the above $F = 3$ level, relative to the $J = 5/2$, $F = 2$ 
level (by $\sim 50$~\%; \cite{guibert78}) and could result in the inversion 
(and anti-inversion) of the corresponding $J = 5/2$ satellite lines (at 
rest frequencies of $\sim 6049$~MHz and $6016$~MHz). A detection of these 
lines  would allow one to independently constrain changes in all three constants 
$\alpha$, $y$ and $g_p$ \cite{kanekar04}; these observations are presently being carried out. 

In summary, we report the first detection of the 18cm OH satellite lines 
at cosmological distances. The 1720~MHz line is found in emission, with a 
luminosity larger by more than an order of magnitude than any other known 
1720~MHz maser; the 1612~MHz line is in absorption, as are the other ground 
state OH lines. The satellite line profiles are found to be conjugate, in 
that the sum of their optical depths is consistent with noise; this implies 
that the two lines arise in precisely the same gas. A comparison between the 
satellite line redshifts then yields constraints on $G \equiv g_p 
[ \alpha^2/y]^{1.849}$, a combination of the fine structure constant $\alpha$, 
the electron-proton mass ratio $y \equiv m_e/m_p$ and the proton gyromagnetic ratio 
$g_p$. We find $\Delta G/G = 2.2 \pm 3.8 \times 10^{-5}$, consistent with the 
different constants remaining unchanged from $z \sim 0.247$ (a lookback time of 
$\sim 2.9$~Gyr) to the present epoch. If $y$ and $g_p$ are assumed to remain 
constant, we obtain $\dal = 0.6 \pm  1.0 \times 10^{-5}$, over the redshift 
range $0 < z < 0.247$. 

\begin{acknowledgments}
We thank Rene Vermeulen for much help with the planning and scheduling of the
WSRT observations. The WSRT is operated by the ASTRON (Netherlands Foundation
for Research in Astronomy) with support from the
Netherlands Foundation for Scientific Research (NWO). We also thank the staff
of the GMRT that made these observations possible. GMRT is run by the National
Centre for Radio Astrophysics of the Tata Institute of Fundamental Research.
\end{acknowledgments}

\bibliography{ms}

\end{document}